\documentclass[a4paper,twoside]{article}

\usepackage{epsfig}
\usepackage{caption}
\usepackage{subcaption}
\usepackage{calc}
\usepackage{amssymb}
\usepackage{amstext}
\usepackage{amsmath}
\usepackage{amsthm}
\usepackage{multicol}
\usepackage{pslatex}
\usepackage{apalike}
\usepackage{url}
\usepackage{tcolorbox}
\usepackage{xcolor}

\usepackage{array}
\newcolumntype{L}[1]{>{\raggedright\let\newline\\\arraybackslash\hspace{0pt}}m{#1}}
\newcolumntype{C}[1]{>{\centering\let\newline\\\arraybackslash\hspace{0pt}}m{#1}}
\newcolumntype{R}[1]{>{\raggedleft\let\newline\\\arraybackslash\hspace{0pt}}m{#1}}

\usepackage{SCITEPRESS}     
\usepackage{color}

\newcommand{\al}{\textit{et al.~}}

\begin{document}

\title{Internet of Things Architectures: A Comparative Study}

\author{\authorname{Marcela G. dos Santos\sup{1}\sup{a}, Darine Ameyed\sup{2}\sup{b}, Fabio Petrillo\sup{1}\sup{c}, Fehmi Jaafar\sup{3}\sup{d}, Mohamed Cheriet\sup{2}\sup{e}}
\affiliation{\sup{1}Département de Informatique et mathématique, Université du Québec à Chicoutimi,  Chicoutimi, Québec, Canada}
\affiliation{\sup{2}École de Technologie Supérieure, University of Quebec, Montreal, Québec, Canada}
\affiliation{\sup{3}Computer Research Institute of Montréal, Montreal, Québec, Canada}
\email{\sup{a}marcela.santos1@uqac.c, \sup{b}fabio@petrillo.com, \sup{c,e}{darine.ameyed.1, mohamed.cheriet}@etsmtl.ca, \sup{d}fehmi.jaafar@crim.ca}
}


\abstract{Over the past two decades, the Internet of Things (IoT) has become an underlying concept to a variety of solutions and technologies that it is now hardly possible to enumerate and describe all of them. The concept behind the Internet of Things is as powerful as it is complex, and for the components in the IoT solution to mesh together perfectly, they all have to be part of a well-thought-out structure. That is where understanding the IoT architecture becomes paramount. Because of the vast domain of IoT, there is no single consensus on IoT architecture. Different researchers and organizations proposed different architectures under a variety of classifications, mainly: conceptual, standard and, industrial or commercial adoption. It is indispensable to make a systematic analysis of IoT architecture to be able to compare the industrial proposals and identify their similarities and their differences. 
In this work, we summarize information about seven IoT industrial architectures in order to propose an approach that make possible a comparative analysis between different IoT architectures. This work presents two main contributions: (i) an approach for analyzing and comparing IoT architectures using Layer-Model; (ii) a comparative study of seven industrial IoT architectures.}
\onecolumn \maketitle \normalsize \setcounter{footnote}{0} \vfill

\section{\uppercase{Introduction}}
\label{sec:introduction}

\noindent The Internet of Things (IoT) marketplace is growing spectacularly in the last few years. Indeed, Huawei expectation for the number of devices connected is 100 billion by 2025. Besides that, the impact of Internet of Things Market on the global economy is enormous. McKinsey Global Institute expected this impact to be around 10 trillion US dollars by 2025 \cite{AHMED2019}. 

Nevertheless, there are many challenges in developing IoT applications: the lack of general guidelines or frameworks that handle low level communication and simplify high level implementation,  the using of multiple programming languages to implement IoT applications, the diversity of the communication protocols, and the high complexity of distributed computing \cite{AMMAR2018}.

Several researchers pointed out the using of IoT Architecture as a tool to clarify the complexity of the IoT solutions and to provide a better comprehension of the issues that may threaten them \cite{ALSHOHOUMI2019}. 

Concretely, Alshohoumi \al \cite{ALSHOHOUMI2019} analyzed 148 studies and identified sixteen different IoT architectures. Moreover, Pratap Singh \al \cite{SINGH2020} affirmed that there is in general three different ways to classify the IoT architectures:  domain-specific architectures, layer-specific architectures, and industrial or commercial defined architectures.

\begin{figure}[ht]
\begin{subfigure}[t]{3.2in}
			\includegraphics[width=1\linewidth]{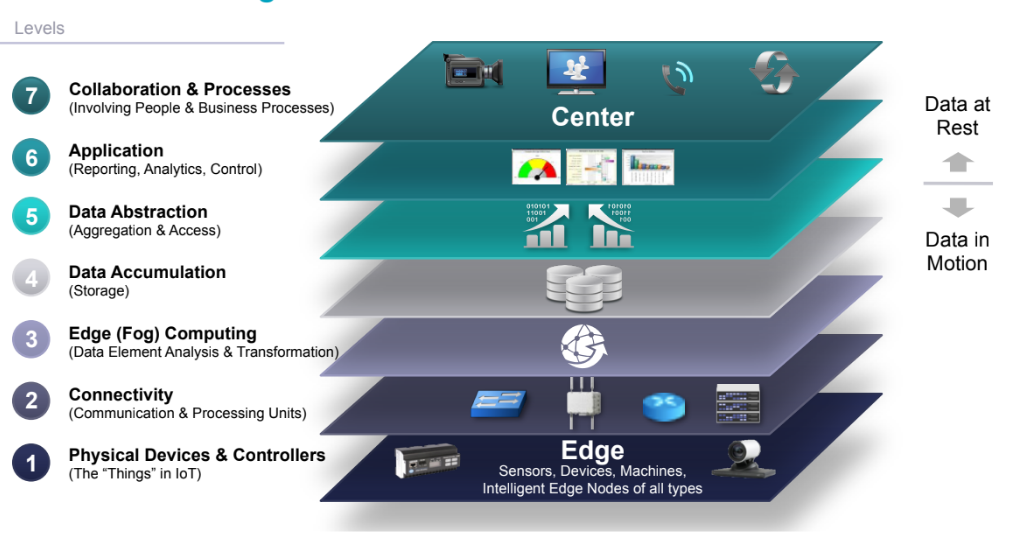}
		\caption{Cisco Architecture \cite{CISCO2014}}\label{fig:ciscoreference}			\end{subfigure}
	\par\bigskip
	\begin{subfigure}[t]{3.2in}
			\includegraphics[width=1\linewidth]{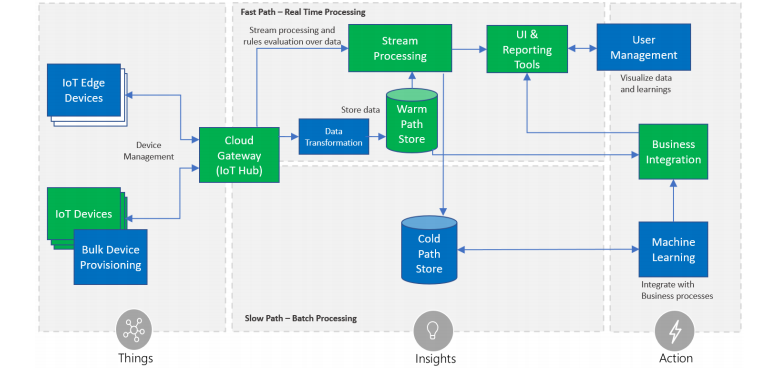}
		\caption{Microsoft Architecture \cite{AZURE2019}}\label{fig:microsoftreference}
	\end{subfigure}
	\par\bigskip
	\caption{Examples of IoT Architectures}\label{fig:examplesiotarch}
\end{figure}

In the industrial context, we noticed that there is a variety of IoT architecture used and presented (Figure \ref{fig:examplesiotarch}). Thus, it is indispensable to make a systematic analysis of IoT architecture based on a reference model to be able to compare the industrial proposals and identify their similarities and their differences.

Thus, in this paper, we selected and analyzed seven industrial architectures in accordance of one of the reference model of IoT architectures, i.e. the Layer-based architecture model.This study presents two main \textbf{contributions}: (i) an approach for comparing IoT architectures in according with a reference IoT architecture and (ii) a comparative study of seven of the main industrial IoT architectures (Intel, Microsoft, Cisco, Google,and IBM, Ericsson, Amazon).

The \textbf{audience} of this study are (i) researchers interested on modeling IoT architectures, and (ii) practitioners interested on comparing industrial IoT architectures.


The remaining paper is organized as follows: Section \ref{sec:BACKGROUND} provides an overview about the IoT Architectures. Section \ref{sec:RELATEDWORK} presents the related work. Section \ref{sec:approach} is the kernel of the work, which provides a approach to analyze and model IoT architectures using Layers-Model. Section \ref{sec:MODELING} presents the modeling of seven industrial architectures using our approach. Section \ref{sec:ANALYZING} performing the analyze of seven industrial architectures and the lessons learned in our study. Section \ref{sec:THREATSTOVALIDITY} explains the threats to validity. Finally, Section \ref{sec:CONCLUSIONS} synthesizes the final remarks and future work.

\section{IoT \uppercase{Layer Model}}
\label{sec:BACKGROUND}

\noindent 



While acknowledging the existence of other IoT architectures classification as conceptual, domain-based and industrial. In our actual study, as mention, we focus on the layered-IoT architectures and industrial architectures.  
Firstly, each layer address points separated before it is integrated and perform as a system. This methodology helps manage the complexity of the system. IoT scenarios have a high level of complexity because of the integration of various kinds of technologies, devices, objects, and services.

The primary studies designed their layered architecture ranged from 3 to 7 layers which are composed by the main building blocks in the IoT platforms, varying from the basic to the end-to-end solutions.

The early IoT model was a three-layer architecture presented by Gubbi \al \cite{GUBBI2013}. Basically, it consists of perception as a ground layer including sensors and actuators as things, cloud as an information processing layer, and application layer that allows users' interaction as the third layer. Extended by adding a business layer to provide the four-layer model \cite{MUCCINI2018}. 

Furthermore, A. Al-Fuqaha \al \cite{AL-FUQUAHA2015} gives the definition of IoT architecture as middleware layer-based and five-layer model. The middleware layer-based, include service composition, service management and object abstraction. Besides, there is the six-layer model adding fog layer or a gateway layer to the five-based model including also edge and hybrid edge-cloud \cite{PAN2018}.  

Finally, the recent proposal for the IoT layered architecture is delivered by Cisco as a seven-layer model. The previous architecture was changed by adding a user and process layer and edge computing layer \cite{CISCO2014}.


\subsection{IoT Layers-Model: 3,5 and 7-Layers Model}
\label{subsec:IOTMODELAYERS}

There are many model Layers Model Architectures described by the literature. For example, in the systematic review about the Internet of Things architecture,  Alshohoumi \al \cite{ALSHOHOUMI2019} noticed, after analyzing 148 studies with 16 mapping architectures, that the most of architectures can be classified as a three, four or five layers. Another survey about IoT architecture is the study conducted by Pratap Singh \cite{SINGH2020} noticed that the model layer classification applied is three, four, five, six, seven layers.

In our current study, we have focus on the 7-Layers architecture model because it is considered the complete model, regarding the complexity of the IoT systems nowadays and the trend in this area.In our current study, we choose to set the focus on the 3, 5, and 7-Layers architecture models for several reasons. Firstly, the 3-Layers model is the most abstract and straightforward model where the majority of research in IoT architecture starts (\cite{KHAN2012} and \cite{WU2010}). Secondly, the 7-Layers is considered the complete model, regarding the complexity of the IoT systems nowadays and the trend in this area. Finally, the 5-Layer model is a model between the other two reference models (3 and 7-Layers model).

\subsubsection{3-Layers Model}
According to Ray \cite{RAY2018} and Lin \al \cite{LIN2017} the trivial architecture for IoT systems is called 3-Layers as it has three layers: perception, network, and application layers  (Figure \ref{fig:3layerssimple}).

On the Perception Layer are the things. The things have sensors that are responsible for taking information, and actuators to interact with the environment. 

The Networking Layer is in charge of connecting the thing with another things, network devices, and servers. 

The last layer on the architecture is the Application Layer that addresses the delivery of the services for the final user, and it is on this layer that is the clouds and servers.

The IoT 3-Layers as an accepted structure but is trivial modeling of the IoT ecosystem  \cite{ALSHOHOUMI2019}.One point important for 3-Layer architecture is the fact that there is not a layer for Business.

\subsubsection{5-Layers Model}

Researchers proposed a new architecture to solve the issues summarized in the 3-Layers architecture. The 5-Layers (Figure \ref{fig:5layerssimple}) is an extension of 3-Layers with the introduction of Processing and Business Layers(\cite{WU2010} and \cite{SETHI2017}).

The characteristics and goals of the Perception and Application Layers are the same as the 3-Layers architecture. The Transport Layer is responsible for transferring the data from things to the Processing Layer in both ways.

The Processing Layer works as a middleware in the 5-Layers architecture; the role of this layer is to store, analyze, and process the information of objects received from the transport layer. With the Processing Layer, various technologies can be applied, for example, database, cloud computing, and big data processing modules. 

Finally, the Business Layer is the layer that threatens all  IoT system management; it includes the application, business, and profit models and user privacy. In the 5-Layers, it is considered data storage and processing, but neither security and privacy are discussed. 

\subsubsection{7-Layers Model}
\begin{figure}[h]
\begin{subfigure}[t]{.3\linewidth}
    \centering\includegraphics[width=.9\linewidth]{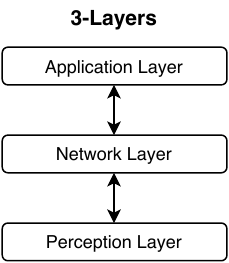}
    \caption{}
    \label{fig:3layerssimple}
  \end{subfigure}
  \begin{subfigure}[t]{.3\linewidth}
    \centering\includegraphics[width=.9\linewidth]{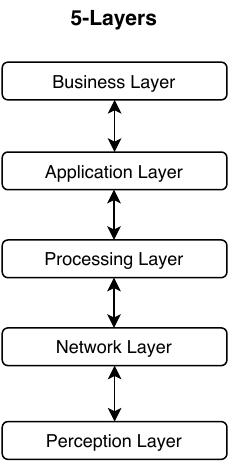}
    \caption{}
    \label{fig:5layerssimple}
  \end{subfigure}
  \begin{subfigure}[t]{.3\linewidth}
    \centering\includegraphics[width=.9\linewidth]{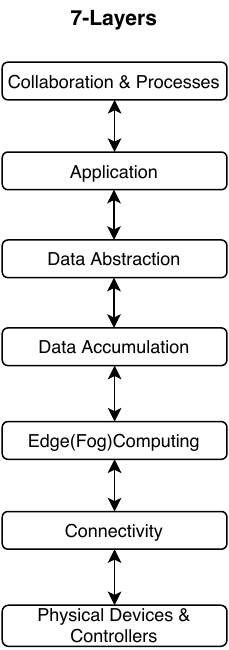}
    \caption{}
    \label{fig:7layerssimple}
 \end{subfigure}
\caption{IoT Layers Architectures}
\label{iotlayersarchsimple}
\end{figure}

The 7-Layers (Figure \ref{fig:7layerssimple}) used in our current study was proposed in the Internet of Things World Forum (IoTWF) \cite{PISCHING2018}. Following the description for each layer.
\begin{itemize}
    \item The first layer is where a variety of devices, sensors, and controllers that enable their interconnection are situated.
    \item The second layer is responsible for making all connections and data transfers in the IoT system. Thus, this layer specifies the communication protocols.
    \item The third layer, the Edge/Fog Computing layer, is where the data analysis and data transformation, is performed \cite{TZAFESTAS2018}.
    \item The fourth layer, the Data Accumulation, leads with the storage data and guarantee that the data is moving correctly. 
    \item The fifth is where the data is prepared to be analyzed using the data mining techniques or the data implementation of machine learning \cite{TZAFESTAS2018}. The last two layers are Application and Collaboration \& Process. 
    \item The Application Layer is where the users can use the information about the environment that is taken by the things. 
    \item Finally, the seventh layer represents the actors that use the data to make a decision based on the data extracted on the IoT ecosystem \cite{PISCHING2018}.
\end{itemize}

\section{\uppercase{Related Work}}
\label{sec:RELATEDWORK}
\noindent A systematic review of existing IoT architectures was performed by Alshohoumi \al \cite{ALSHOHOUMI2019} on a set of 144 studies from 2008 to 2018. It is a review of IoT architectures in terms of architecture classification (the number of layers), limitations in each architecture, and considerations of different aspects or features in each layer. Our study differs from the review perform by Alshohoumi \al \cite{AMMAR2018} in the following terms: (i) we investigate the architecture proposed by IoT providers, (ii) we apply an approach to model and analyze the architecture following for IoT providers.  

Ammar \al \cite{AMMAR2018} presents a survey on security of IoT frameworks; a total of 8 frameworks are considered. Their study had as a goal:  clarifying the state of the art IoT platforms and identifying the trends of current designs of such platforms, providing a high-level comparison between the different architectures of the various frameworks. Our study differs from their because, in our research, we model the IoT providers using our approach and before performing a comparison. 

Pratap Singh \al \cite{SINGH2020} presents an analysis of Layer-Specific, Domain-Specific, and Industry defined IoT architectures.  The main contribution is a summarizing of state of the art of IoT with a comparison of the industry-defined architectures. The work by Pratap Singh \al and our study can be considered as complementary each of them analyzes IoT architectures. 
Still, our study proposes an approach for modeling  IoT architectures and also compares seven industrial IoT architectures (Intel, Microsoft, Cisco, Google, and IBM).

\section{\uppercase{Approach for modeling and analyzing of} IoT \uppercase{architectures}}
\label{sec:approach}
\noindent In this section, we discuss our approach for modeling and analyzing IoT architectures. The evaluation approach which we propose is built upon the referential architecture described in the section \ref{sec:BACKGROUND}. First, we discuss the construction, and then we summarize the steps that composed our approach.


We start analyzing the documentation that IoT providers make available publicly. These data are available on websites or in white papers (\cite{INTEL2015}, \cite{AZURE2019}, \cite{GCP2019}, \cite{IBM2019}, \cite{CISCO2014}, \cite{AMAZON2016}, \cite{ERICSSON2017}).

To analyze this information systematically, we summarize essential aspects for each layer in the IoT architecture by the provider. We read the documentation and conduct an extraction of some elements for each layer: input, output, activities performed, and the principal objective. It is important to note that some aspects have different terminology or functionalities in each layer for each provider. 

Based on the extracted data for each layer in each provider (input, output, activities, and main objective) for the 7-Layers Model, we classified the layers in the industrial architecture according the layer in the referential model. For example, for the architecture following by Microsoft, we compare each layer with the reference model using the data extraction for Microsoft IoT white paper \cite{AZURE2019}, and we could classified each layer in the Microsoft proposal in most similar layer in the 7-Layers model.

It is important to highlight there is some layer that are  not described neither in the documentation nor in the figure represented the architecture. In this cases, we decided to represent the layer using dotted line to demonstrated that this layer does not have description in the available documentation,

In our approach for each layer in the IoT architecture in question, we extracted data to answer the following set of questions using the information made available for the IoT provider: 
        \begin{itemize}
            \item What is(are) the input(s) for the layer?
            \item What is(are) the output(s) for the layer?
            \item What does the layer perform the activities?
            \item What is the principal objective of the layer?
        \end{itemize}

\section{\uppercase{Modeling seven IoT Providers using our approach}}
\label{sec:MODELING}
\noindent In this section, we apply the approach described in Section \ref{sec:approach} for seven leading industrial actors: Intel, Microsoft, Cisco, Google, IBM, Ericsson, and Amazon.

Our inclusion criteria for the IoT architectures were selecting architectures that were studied at least twice by Ammar \al \cite{AMMAR2018}, Lueth \cite{IOTANALYTICS2015} and Asemani \al \cite{ASEMANI2019}. We used these studies as a guideline to select the IoT architectures because of the following reasons: (i) these researchers perform a study similar than our study, (ii) these studies analyzed deeply the IoT architecture and  (iii) these studies analyzed the leading IoT companies around the world.

\subsection{Intel Architecture}

Intel IoT Platform Reference Architecture is a system architecture specification (SAS) to connect any products and services to the cloud \cite{INTEL2015}. There are two versions of this IoT Architecture. Firstly, version 1.0 is an architecture to connect the unconnected things, and the version is the Intel SAS version 2.0 that brings a specification about the integration of a variety of devices with intelligence and connectivity integrated \cite{BREIVOLD2017}.

In our study, we classified the Version 2.0 because of it being a reference architecture that facilitated the convergence of operational technology and information technology. Additionally, it is a future-looking reference architecture. 

\begin{figure}[ht]
    \includegraphics[width=\linewidth]{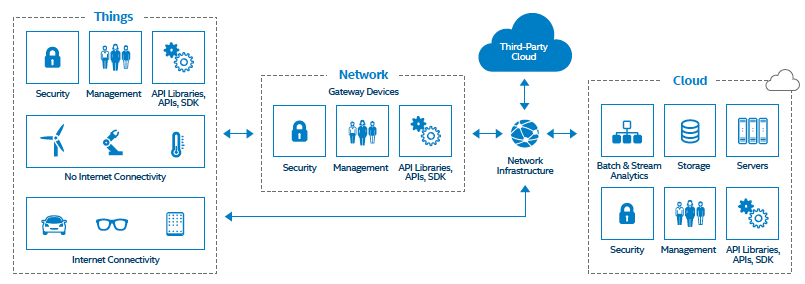}
    \caption{Intel End-to-End IoT Solution form Things to Network to Cloud. \cite{INTEL2015}}
    \label{fig:inteloverview}
\end{figure}

According to the \cite{INTEL2015}, the three main components are: things, network and cloud (Figure \ref{fig:inteloverview}). The Things component is not represent in the Layered Architecture but as we notice in the documentation available, Intel defined a end-to-end solution for connecting nearly any type of device to the cloud.

\begin{figure*}[ht]
\centering
   \includegraphics[width=.65\linewidth]{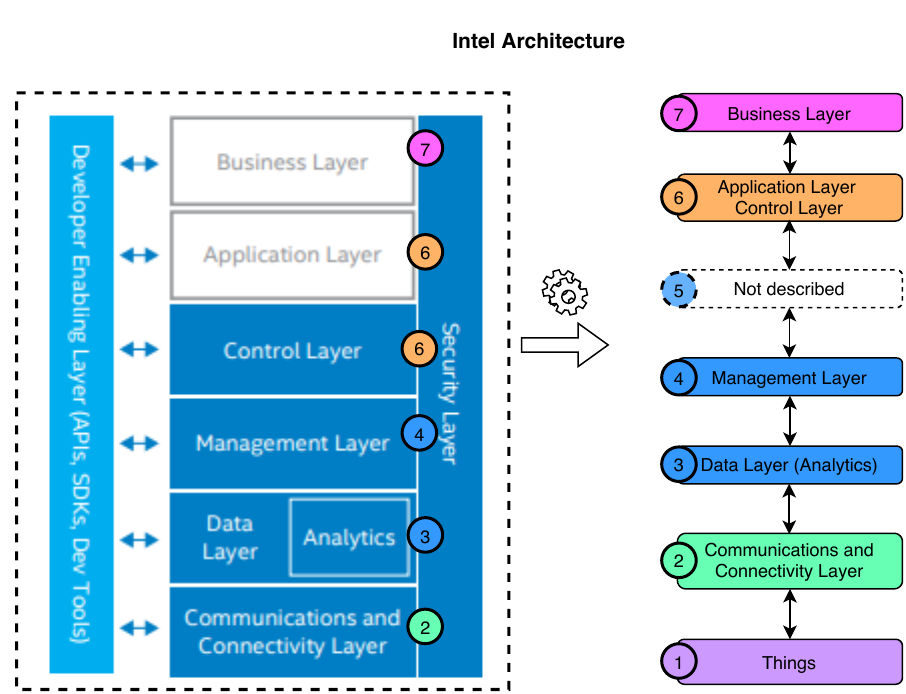}
  \caption{Analyze of Intel IoT Architecture.}
 \label{fig:intelml}
\end{figure*}

In Figure \ref{fig:intelml} more specifically in the figure more in the right and dotted, we have the architecture proposes by Intel \cite{INTEL2015}. The first group of layers (in dark blue) have layers that are the major run-time layers. The Communication and Connectivity layer is situated on the bottom of the architecture, and are responsible for enabling multi-protocol data communication between not only devices and the edge but also between endpoint devices/gateways, the network, and the data center \cite{INTEL2015}.

The second layer is the Data layer with Analytics, whose role is providing customer value. It is achieved using valuable insights generated by data analytics and improved closed-loop control systems. On the Intel IoT reference architecture, this need is addressed by allowing analytics to be distributed across the cloud, gateways, and smart endpoint devices \cite{BREIVOLD2017}. 

The Management layer is the next layer in which the primordial role is for realizing automated discovery and provisioning of endpoint devices. Intel recommends Device Cloud product to perform the manageability functions \cite{BREIVOLD2017}. And to complete the dark blue group, the Control layer that provides a way to separate the management layer into a management plane and control plane, with policy and control objects and APIs.

Now, the second group is composed of user layers. The first is the Application layer that is used by the Business Layer to access the other layers in the Intel IoT architecture. Besides these layers, there is a vertical security layer. The role of the security layer is to address protection and security across all tiers.

We could summarize some aspect about Intel architecture, applying our approach. The first aspect important in the Intel architecture is that there is a Business Layer, what shows that for Intel is important to have clearly the IoT economic impact. And, in Intel architecture, the Data Abstraction layer (Figure \ref{fig:intelml}) is not described neither in the documentation nor in the layer architecture


\subsection{Microsoft Architecture}

The design of the Microsoft IoT architecture is based in three layers: Thing, Insight, and Actions. There are a set of subsystems in all layers, and it is essential to highlight that Edge Devices, Data Transformation, Machine Learning, and User Management are optional subsystems \cite{PROTOCOLGATEWAY}.

The architecture proposes by Microsoft (Figure \ref{fig:microsoftreference}) the first group of subsystems are in the layer ``Things" and are IoT Devices, IoT Edge Devices, and Device Provisioning. 

The IoT devices are devices that need to register in a way secure with the cloud to send and receive data. The second group is on the layer ``Insights" and consists of Cloud Gateway, Data Transformation, UI and Reporting Tools, Stream Processing, Warm Path Store, Cold Path Store. 

In the case of the devices or field gateways that are not able to use the standard protocols used by IoT Hub, adaptation is necessary, and Azure IoT protocol gateway can be used \cite{PROTOCOLGATEWAY}. After that, we have in the Microsoft IoT architecture the Stream Processing that is responsible for consuming the massive streams of data records and evaluate rules for those streams.The third layer is the ``Action" layer. On this layer, we have the Business integration, which essential role is performing actions based on the telemetry data during stream processing. Also, in the ``Action" layer, we have the User Management subsystem whose goal is restricting the user or user groups action on the devices. And finally, we have the Machine Learning subsystem that is responsible for performing predict algorithms using the telemetry data. This prediction can be applied in predictive maintenance, for example.

We apply our approach and could summarize similarities and divergences between the layers in the Microsoft architecture and the layers in the 7-Layers. We could find similarities between the layers IoT Edge Devices and Cloud Gateway and the Transport layer.

The Cloud Gateway and Data Transformation modules in the architecture follows by Microsoft have similarities with the layer Edge(Fog)computing in the referential model 7-Layers. The Data Accumulation is represented by the Warm Path Store and Cold Path Store. For the Data Abstraction in the referential model, we have the Stream Processing in the Microsoft architecture.

For the Application layer in the 7-Layers model, we could find similarities with the modules UI and Report Tools, User Management and Machine Learning. Finally, the activities that composed the Business layer are performed by the module Business Integration.


\begin{figure*}[ht]
    \includegraphics[width=\linewidth]{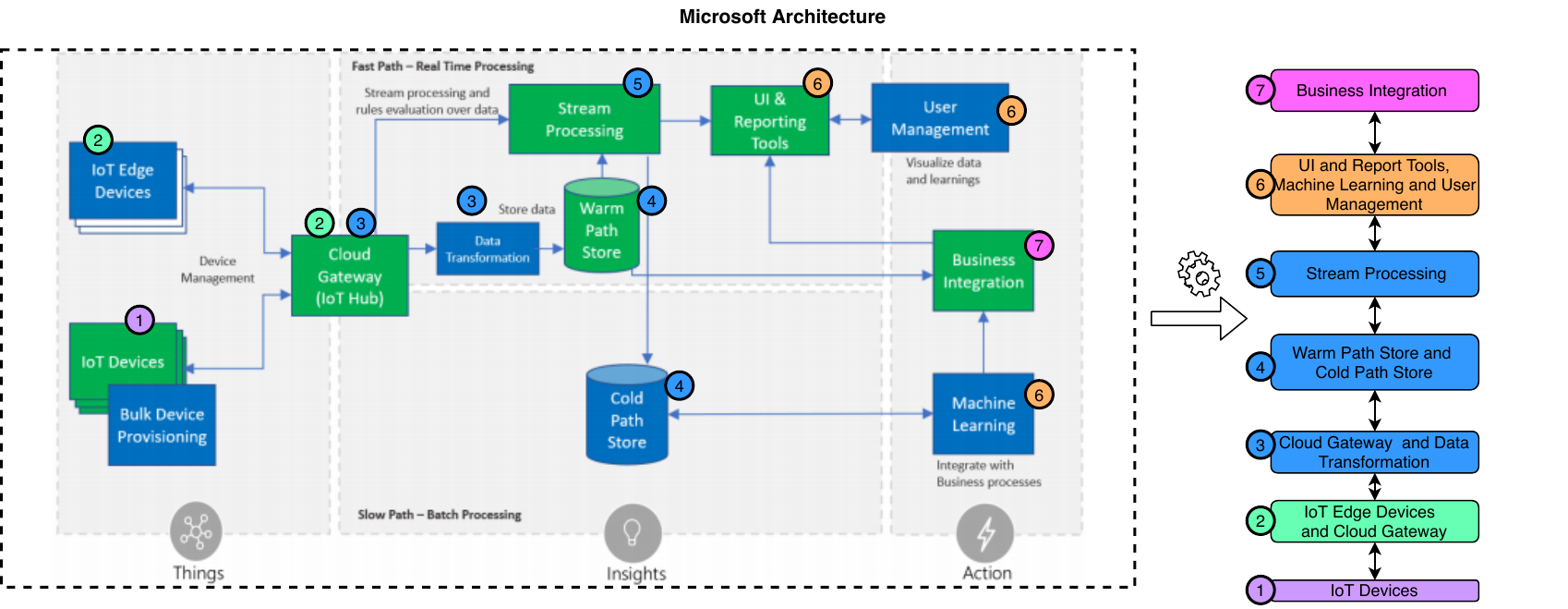}
    \caption{Analyze of Microsoft IoT Architecture}
   \label{fig:microsoftml}
 \end{figure*}

\subsection{Cisco Architecture}

\begin{figure*}[ht]
    \includegraphics[width=\linewidth]{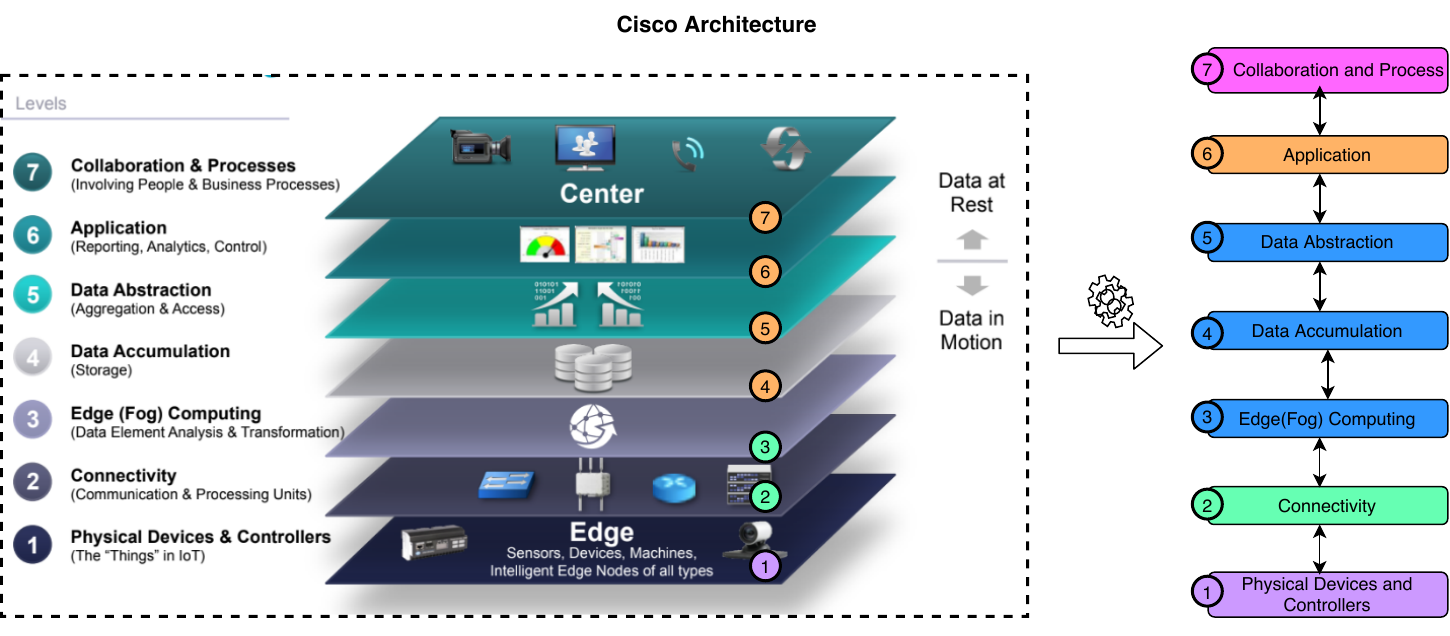}
    \caption{Analyze of  Cisco IoT Architecture.}
    \label{fig:ciscoml}
\end{figure*}

Cisco proposal(Figure \ref{fig:ciscoml})is a multi-level reference that is composed of seven layers in the dotted area shows \cite{CISCO2014}. 

In level 1 is the Things on the Internet of Things. As an example, we can have endpoint devices that may send and receive information. Level 2 is focused on communication and connectivity. The connectivity includes the data transmission between devices and the network, across networks and between the network and low-level information processing. The activities in level 3 are focused on reorganizing the data that is produced by the things in the level 1 into information that is convenient for level 4 (storage and higher processing). The priority of level 4 is guaranteeing that the data is moving precisely. In other words, the data is moving at the rate and organization defined by the devices that generated the data in level 1. Data Accumulation level will make network data turn into data usable for the application. The idea is to convert data-in-motion to data-at-rest, converting data network packets to relational database tables.

After the data accumulation is necessary to render the data and store this information in a way to facilitate the development of application more simple and with higher performance, this is performing by the level 5.Level 6 is the application level, where all the data that is generated is interpreted using various types of an application. The collaboration \& process level execute the applications with the specific needs. With the apps (level 6), people have access to the right data at the right time so that they can do the right thing \cite{CISCO2014}.

We applying our approach and find just similarities between the layers in the Cisco architecture and the layers in the 7-Layers. For us, the analysis of Cisco was proof that our approach is a valuable tool to be used to model and analyze IoT architectures.


\subsection{Google Architecture}

Google Cloud Platform (GCP) is a complete set of tools to connect, process, store, and analyze data both at the edge and in the cloud \cite{GCP2019}. To GCP, the IoT solution consists of three essential components, the device, gateway, and cloud. The element that is directly related to the world is the device. For this reference architecture, the device can be hardware and software and might be directly or indirectly connected to the Internet. Besides that, devices can communicate with each other via a network.

The gateway is responsible for guaranteeing the devices that are not directly connected to the Internet to reach cloud services. Another aspect of the gateway is that it processes data on behalf of a group or cluster of devices. The devices collect data; the gateway sends this data to Cloud Platform, the third part of the architecture. On the cloud platform, the data are processed and combined with other data that were sent for different devices. 

\begin{figure*}[ht]
    \includegraphics[width=\linewidth]{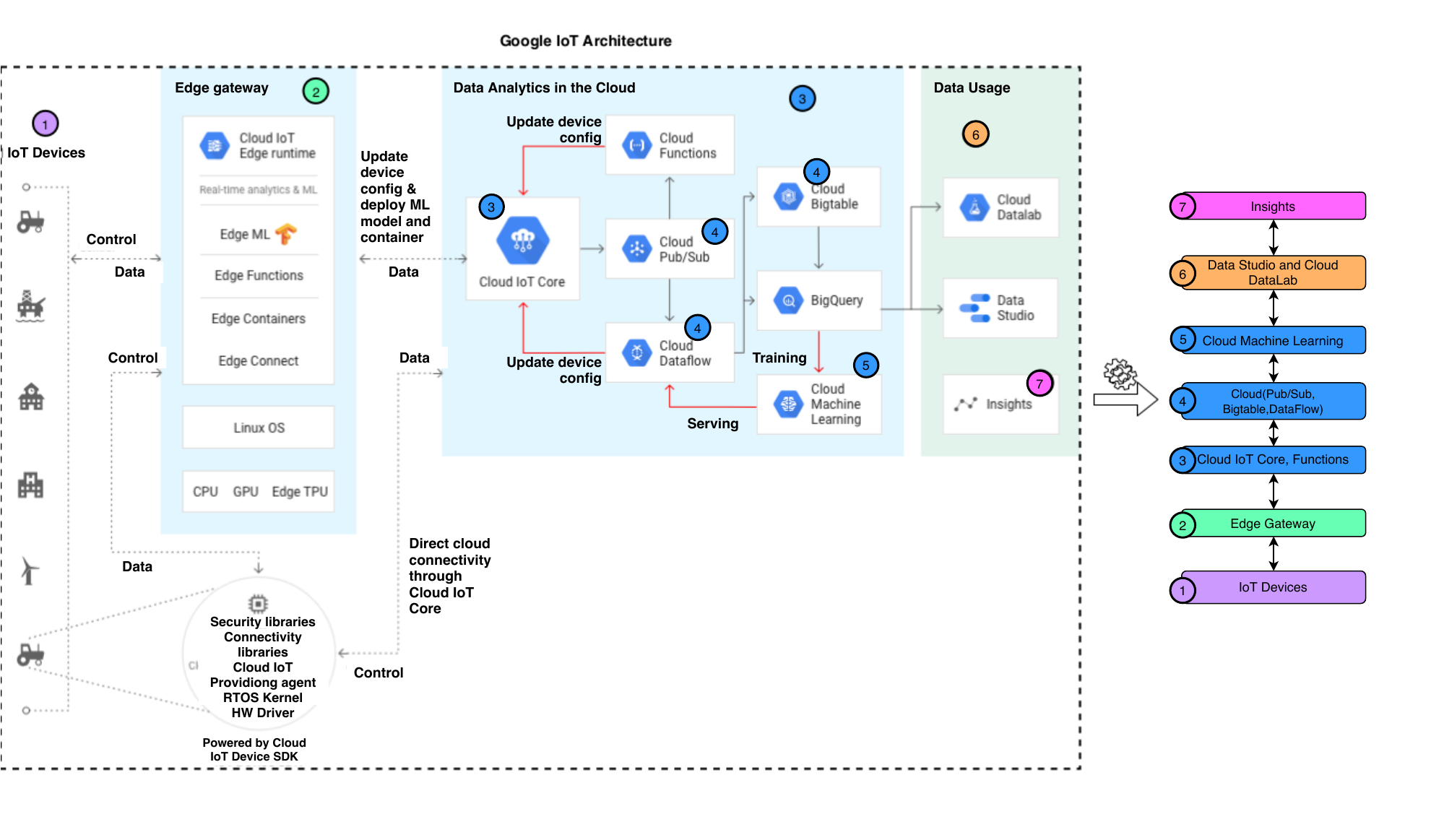}
    \caption{Analyze of Google IoT Architecture.}
    \label{fig:googleml}
\end{figure*}

We analyzed the similarities and divergence between Google architecture and our referential model (Figure \ref{fig:googleml}). In our analysis, we concluded that there is a high level of linkage between the internal modules in the Data Analytics in the Cloud. For example, there are three modules in the Data Analytics in the Cloud that can be classified as Data Accumulation in the 7-Layer model Although Google uses only four elements to present the IoT architecture following in its solutions, it was necessary to ''open'' the modules to understand deeply what the activities performed and how we could classify using our approach.

\subsection{IBM Architecture}

\begin{figure*}[ht]
    \includegraphics[width=\linewidth]{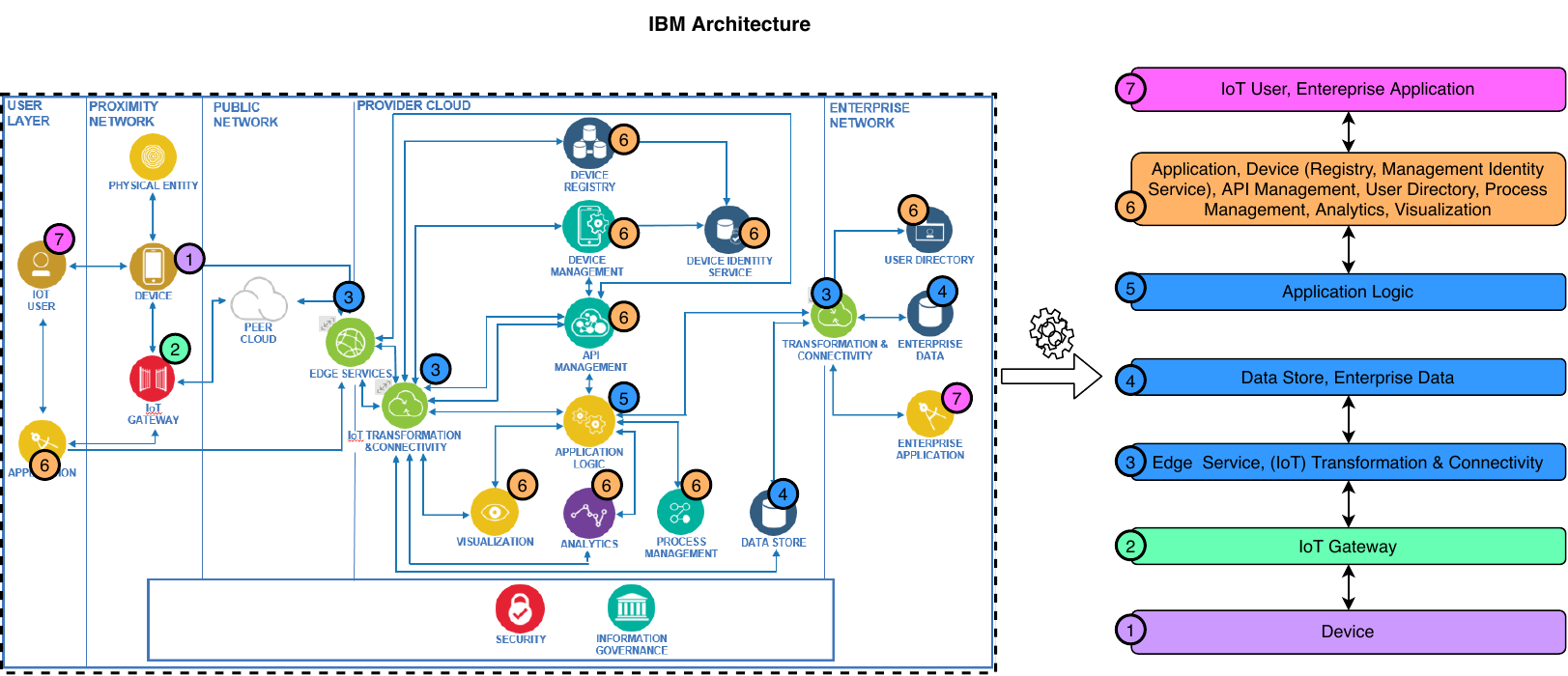}
    \caption{Analyze of IBM IoT Architecture.}
    \label{fig:ibmml}
\end{figure*}

The IBM (Figure \ref{fig:ibmml}) proposal for IoT architecture is composed by the following layers User Layers, Proximity Network, Public Network, Provider Cloud, and Enterprise Network. The main objective of IBM IoT architecture is to perform a connection to IoT devices and quickly build scalable apps and visualization dashboards to gain insights from IoT data. IBM IoT architecture uses IBM Cloud IoT, data, and AI services to achieve the main goal \cite{IBM2019}.

There are five layers called the User Layer, Proximity Network, Provider Cloud, and Enterprise Network. We could find similarities between the layers in the IBM architecture and the 7-Layers model with the application of our approach. But it is essential to highlight that we made the analysis using the internal modules insight the layers: User Layer, Proximity Network, Provider Cloud, and Enterprise Network.

The sensors (water leak detection, water flow, temperature, etc.) and actuators (automatic water shutoff valves) are in the Proximity Network; this layer can be at home for example, if we talk about IoT platforms for Smart Homes. They are attached to the device maker's cloud service. 

In this scenario, we can see an IoT application, the Smart homes. The connected devices let insurance companies improve service for their policyholders. Besides that, with data collected is possible to provide insight about risks that can happen in the home. For example,  leak-detection sensors can monitor for water leaks. With a correct analyze, valves can be trigged, and the IoT platform can help protect the home from resulting damage. 

The Device performs activities comparable with the Perception Layer. The IoT Gateway is the Transport Layer in the IBM architecture. 

In the layer numbering with three, we have more than one module in the industrial architecture (IBM) performing as one layer in the referential model. Edge Service, IoT Transformation \& Connectivity, and Transformation \& Connectivity compos the third layer in the IBM architecture as is comparable with the Edge(Fog) Computing in the 7-Layers model.

The modules Data Store and Enterprise Data numbering perform activities similarities with the Data Accumulation. Application logic (IBM architecture) is the Data Abstraction(7-Layers model).

All device modules (registry, management, and identity service), API Management, User Directory, Process Management, Analytics, and Visualization compos the sixth layer in the IBM architecture and is comparable with the Application Layer in the 7-Layers model.

Finally, IoT User and Enterprise Application are the seventh layer in the IBM architecture and are comparable with the Collaboration and Processes in the 7-Layers.


\subsection{Ericsson Architecture}
\label{subsec:ericsson}
\begin{figure*}[h]
    \includegraphics[width=\linewidth]{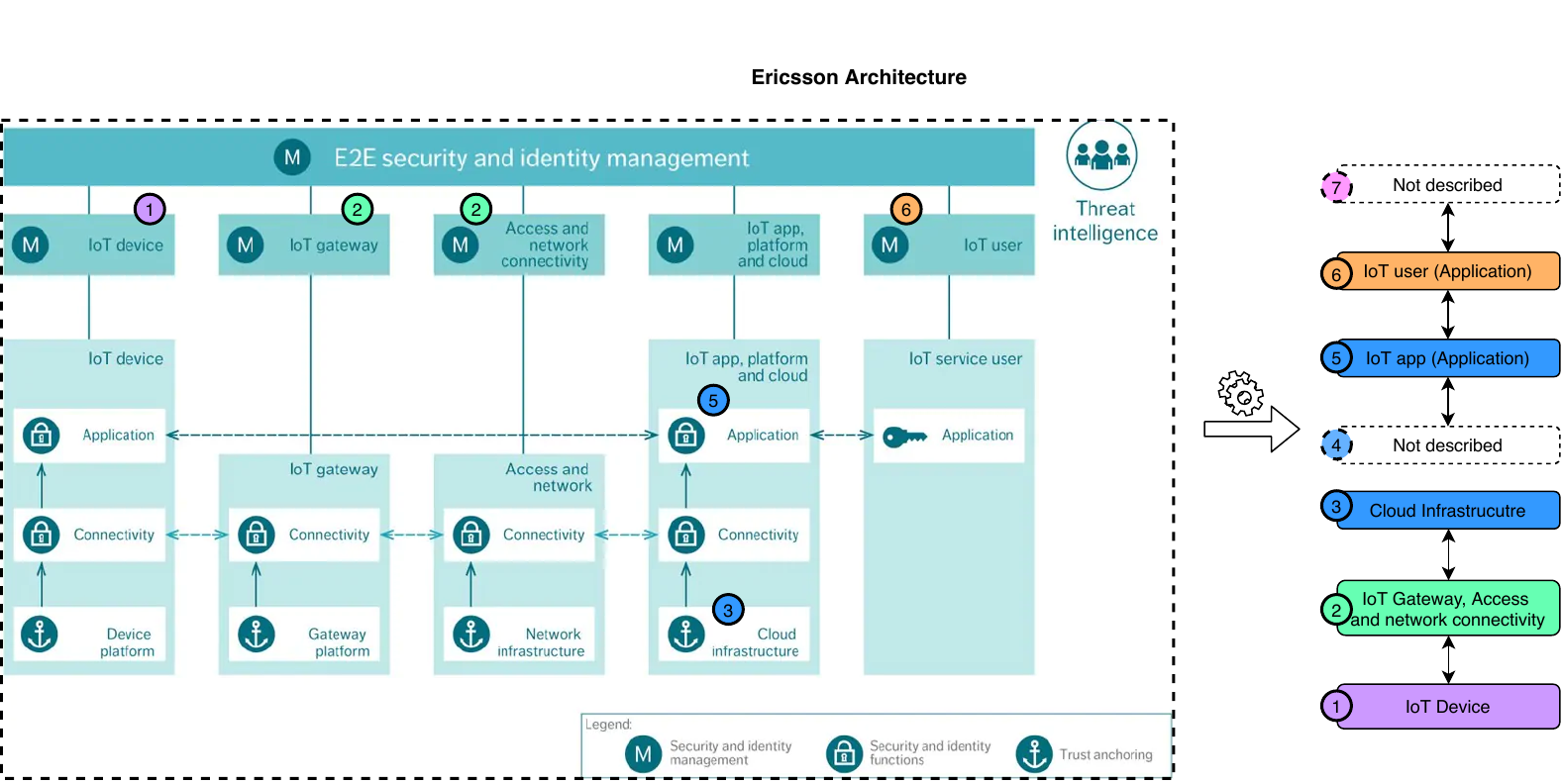}
    \caption{Analyze of Ericsson Architecture}
    \label{fig:ericssonml}
\end{figure*}

Ericsson's IoT architecture proposal (Figure \ref{fig:ericssonml}) is a security solution that provides continuous monitoring of threats, vulnerabilities, risks, and compliance, along with automated remediation \cite{ERICSSON2017}. 

We analyze the architecture proposal by Ericsson using our methodology. We classified the IoT devices as the Perception layer, the IoT Gateway, Access, and network connectivity as a Network layer. In the layer IoT app, platform \& cloud, we could find similarities between the activities performed in the module Cloud Infrastructure with Edge(Fog)Computing, and Application with Data Abstraction, we do not find the description about the Data Accumulation. IoT User as Application Layer.We do not find represent the Business Layer.

The Business Layer and Data Accumulation in the Ericsson architecture is not represented but all the other layers have similarities with the layer in the 7-Layers model.


\subsection{Amazon Architecture}

The main objective of the AWS IoT is providing secure, bi-directional communication between Internet-connected devices such as sensors, actuators, embedded micro-controllers, or smart appliances, and the AWS Cloud \cite{AMAZON2016}. 

\begin{figure*}[h]
    \includegraphics[width=\linewidth]{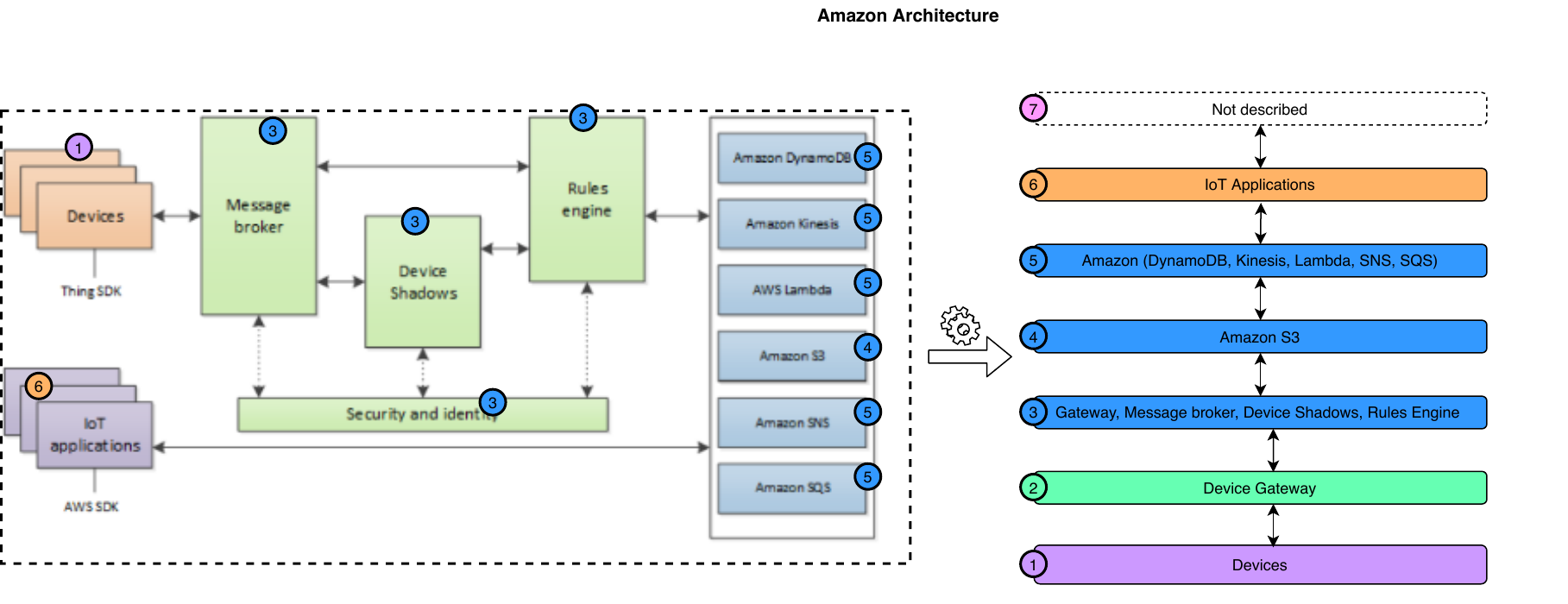}
    \caption{Analyze of Amazon Architecture: the Business Layer is not clearly represented in the architecture documentation(\cite{AMAZON2016}) but all the other layers have similarities with the 5-Layers model.  }
    \label{fig:amazonml}
\end{figure*}

The elements that compose the architecture are Alexa voice service integration for AWS IoT, custom authentication service, device gateway, device provisioning service, device shadow, device shadow service, group registry, jobs service, message broker, registry, Rules engine and security, and integrity service \cite{AMAZON2016}.

We analyze the architecture proposal by Amazon(Figure \ref{fig:amazonml}) using our approach and classified the component Devices as a Perception Layer. The Amazon documentation describes the Gateway component, but it does not show in the architecture. The core of the AWS solution \cite{GUTH2016} is composed of the components numbering by 3, and they have similarities with the Processing in the 7-Layers model. As well as in the Ericsson architecture (Sub-Section \ref{subsec:ericsson}), there is no description of a Business Layer in the Amazon architecture.

\section{\uppercase{Analyzing seven IoT Providers using our approach}}
\label{sec:ANALYZING}
\begin{figure*}[h!]
    \includegraphics[width=\linewidth]{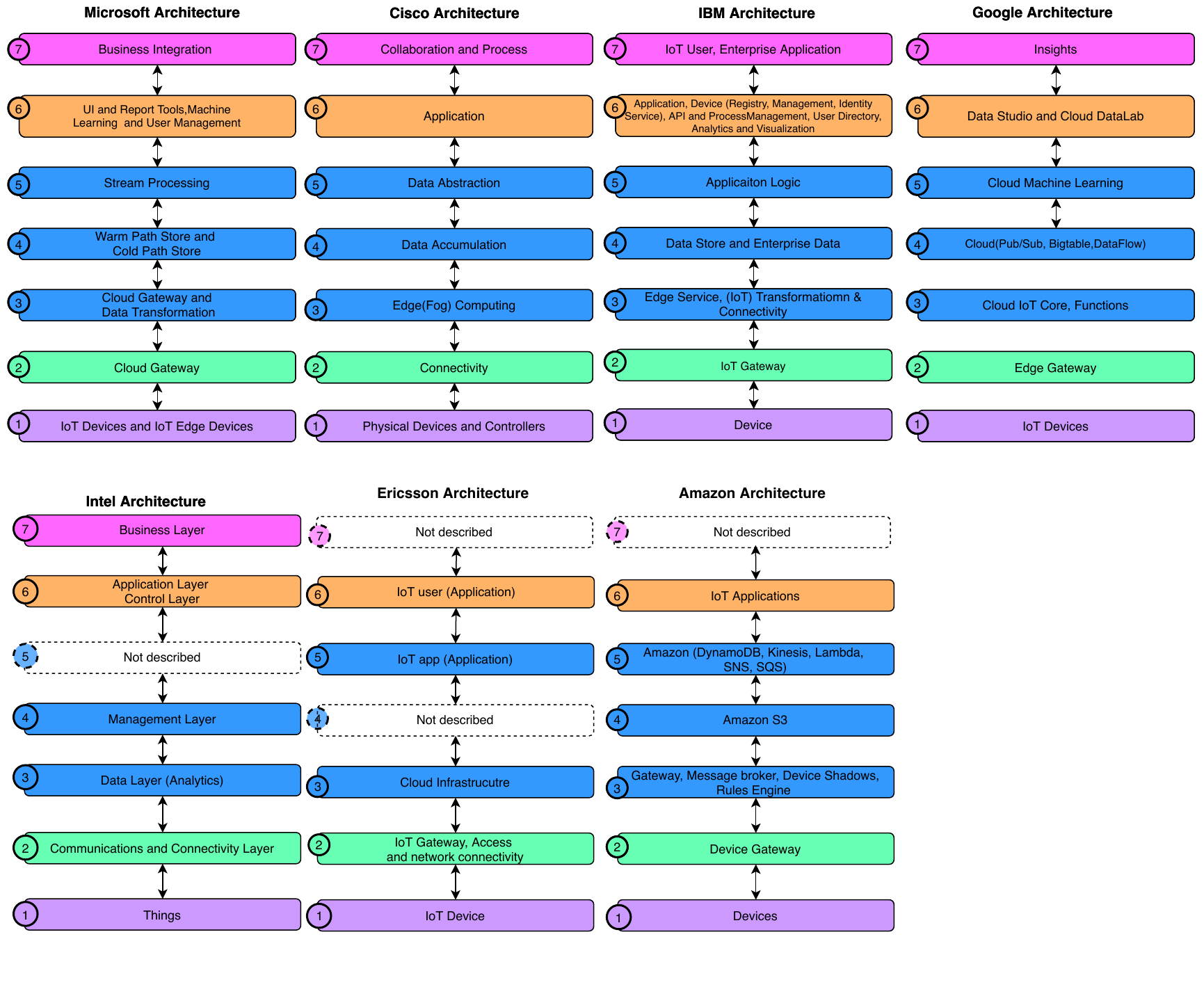}
    \caption{Industrial providers classified as a Layers-Model}
    \label{fig:results}
\end{figure*}

\noindent This section presents a comparative analysis of seven industrial architecture modeled in Layer-Model using our approach (Figure \ref{fig:results}).

Although the fact the seven industrial architectures have terminology different, it was possible to compare the layers providers to the layer-model using the methodology that we proposed. We noticed as far as that the abstraction level concerning the physical layer increases, the terminology became more diverse.

We could classified all the layers in the seven architectures studied into seven layers. In other words, we use 7-layers model to identify the layers in the industrial architecture that can be considered similar. 




For the network layer in the 7-Layers model, we notice that five of the seven industrial have used the term Gateway or similarities to express the layer responsible for performing the communication with and between the perception layer. It is important to clarify the gateway has a specific function in networking, it is also used to describe a class of device that processes data on behalf of a group or cluster of devices \cite{GCP2019}.

Five industrial architectures defined, in the documentation, the Business layer, and three of them use the term Business or similar to denoted the last level of them architectures.  We conclude that for the majority of the companies is important to represent the economic impact of the IoT solutions. Another aspect in the last layer for the industrial architectures and become a recommendation for Ericsson and Amazon is making explicit the Business layer in them architectures. 

Our study could show in a comparative approach the fact that is known in the IoT community: there is no single consensus on IoT architecture. The complexity of IoT architecture is one of the reasons for many proposal architectures.  With decreasing complexity, using a referential model, we could evaluate different architectures.

We learned during the IoT architecture study some major lessons: 
\begin{itemize}
    \item The use of layers to represent architecture is a way more clear to understand the architectures. 
    \item There are some layers in the industrial IoT architectures analyzed that perform the same activities but with different nomenclature.
    \item There are modules in industrial architecture that perform activities inherent to more than one layer in the layer-model architecture.
    \item It is necessary a standard to provide a better comprehension of the issues and threats in IoT.
\end{itemize}

\section{THREATS TO VALIDITY}
\label{sec:THREATSTOVALIDITY}
\noindent During our study, we have counted some threats that we need to address. We work carefully to mitigate eventual issues that could compromise the validity of our results or conclusions. In this section we highlight some of those threats and what mechanism that we applied to address it.

First, the main limitation of this work is the analysis of the IoT architecture by the providers was performing using only the information available in sites and white paper.  However, we opted to perform the analysis only with the data made available for the companies because it is this information that our audience also has to analyze and compare the IoT providers. Besides, we could extract the data, proposes an approach, and compare the architectures.

The second threat to validity is the bias created by the fact that we used one type of referential model, and we chose three models. However, as this, our mapping is preliminary work; consequently, we decide to start the study only with these models and after as a future work performing the same analysis with other models.

\section{CONCLUSION}
\label{sec:CONCLUSIONS}

\noindent 
IoT market is continually growing; as a consequence, the number of IoT architectures has increased too. There is no unique or consensus about the IoT architecture; each architecture has the approach and the interpretation of IoT for the provider. 

In this study, we summarize information about seven IoT industrial architectures. And, we provide an approach that makes possible the modeling of IoT industrial architectures based on Layers-Model, as well as the comparison between IoT architectures.

With our analysis, we could concluded that even though the industrial architectures have represented IoT solutions in a different way with a different terminology, architectures perform the same activities. Besides that, it was clear that some architectures following the seven-layers model, for example, Microsoft and IBM) but others are used the model differently, for example, Intel and Google.

We could conclude not only the need for a standard for IoT architecture and but also the advantages to represent the architectures using layers. With the layers, it was easier to compare and analyze better the industrial IoT architectures.

We aim that our study benefit researchers that performing analysis of IoT architectures and practitioners that need to choose IoT providers. Future work includes (i) a quantitative analysis of IoT architectures studied, (ii) a systematic literature review of Internet of Things architectures, and (iii) systematic analysis about the products that are available by the providers.

\section*{\uppercase{Acknowledgements}}

\noindent To the members of the \textit{SmArtSE Research Team}, Synchromedia Laboratory and Computer Research Institute of Montréal (CRIM) for their support and knowledge sharing.
This work was financed by the Canadian program MITACS and LabVI (http://quartierinnovationmontreal.com/en/open-sky-laboratory-smart-life).

\bibliographystyle{apalike}
{\small
\bibliography{main}}

\end{document}